\documentclass[twocolumn,twoside,slac_two]{revtex4}
\usepackage{graphicx}
\usepackage{fancyhdr}
\begin{document}

\title{Improved limits on sterile neutrino dark matter from full-­sky observations by the Fermi-­GBM}

\author{Shunsaku Horiuchi$^1$, Kenny C.~Y.~Ng$^2$, Jennifer M.~Gaskins$^3$, Miles Smith$^4$, Robert Preece$^5$}
\affiliation{$^1$Center for Neutrino Physics, Department of Physics, Virginia Tech, Blacksburg, VA 24061}
\affiliation{$^2$Center for Cosmology and AstroParticle Physics (CCAPP), Ohio State University, Columbus, OH 43210}
\affiliation{$^3$GRAPPA Institute, University of Amsterdam, 1098 XH Amsterdam, The Netherlands}
\affiliation{$^4$Jet Propulsion Laboratory, Caltech, 4800 Oak Grove Drive, Pasadena, 91109 CA }
\affiliation{$^5$Dept. of Space Science, University of Alabama Huntsville, Huntsville, AL 35805}

\begin{abstract}
For the first time, we use the Gamma-ray Burst Monitor (GBM) on-board the {\it Fermi} satellite to search for sterile neutrino decay lines in the energy range $10$--$25$~keV corresponding to sterile neutrino mass range $20$--$50$~keV. This energy range has been out of reach of traditional X-ray satellites such as {\it Chandra}, {\it Suzaku}, {\it XMM-Newton}, and gamma-ray satellites such as {\it INTEGRAL}. Furthermore, the extremely wide field of view of the GBM opens a large fraction of the Milky Way dark matter halo to be probed. We start with 1601 days worth of GBM data, implement stringent data cuts, and perform two simple line search analyses on the reduced data: in the first, the line flux is limited without background modeling, and in the second, the background is modeled as a power-law. We find no significant excess lines in both our searches. We set new limits on sterile neutrino mixing angles, improving on previous limits by approximately an order of magnitude. Better understanding of detector and astrophysical backgrounds, as well as detector response, can further improve the limit. 
\end{abstract}

\maketitle

\thispagestyle{fancy}


\section{Introduction}

Right-handed neutral fermions (henceforth sterile neutrinos) arise in most implementations of the seesaw mechanism to generate neutrino masses, and yield extremely rich phenomenology (for recent reviews, see, e.g., \citep{Kusenko:2009up,Boyarsky:2009ix}). In particular, in the mass range of 1--100 keV, sterile neutrinos can be produced in sufficient quantities in the Early Universe to be a viable dark matter candidate. If the production proceeds via oscillations with active neutrinos, the momentum distribution of the sterile neutrino results in a warm dark matter candidate \cite{Dodelson:1993je}. However, other production mechanisms have been shown to result in sterile neutrino dark matter that act very similarly to cold dark matter \cite{Shi:1998km,Abazajian:2001nj,Asaka:2006nq,Boyarsky:2008xj,Boyanovsky:2010pw}. Astrophysically, a dark matter sterile neutrino lies in the parameter space to be produced in core-collapse supernova cores \cite{Kusenko:1997sp}, providing a new mechanism for explosion \cite{Hidaka:2006sg}, as well as explaining the origin of strong neutron star kicks \cite{Fuller:2003gy,Barkovich:2004jp,Fryer:2005sz}.

As a viable dark matter candidate, sterile neutrinos are stable on cosmological time scales, but nevertheless they have decay channels that become interesting indirect detection targets for large concentrations of dark matter. The primary decay we target is the radiatively decay channel into an active neutrino and a photon. As the photon carries half of the sterile neutrino mass energy, the photon lies in the X-ray range. The signal is spectrally distinct from most expected astrophysical backgrounds, since the signal line is broadened by the velocity dispersion of the dark matter particles. Coupled with the expected spatial morphology -- spherical and centrally concentrated at the Galactic center -- searches with X-rays can be a very powerful probe \cite{Abazajian:2001vt}.
 
Many searches have been performed in the past, using X-ray satellites such as {\it Chandra}, {\it Suzaku}, and {\it XMM-Newton}; observing a wide range of targets, from galaxy clusters, nearby galaxies, and dwarf satellites of the Milky Way, to the cosmic X-ray background. Most recently, an unexplained X-ray line was detected from a stack of galaxy clusters as well as the Andromeda galaxy \cite{Bulbul:2014sua, Boyarsky:2014jta} (see also followup studies supporting and refuting these initial claims, e.g., \cite{Riemer-Sorensen:2014yda, Jeltema:2014qfa,Boyarsky:2014ska, Malyshev:2014xqa, Anderson:2014tza,Tamura:2014mta,Urban:2014yda}), which can be interpreted as the decay of 7~keV sterile neutrinos \cite{Abazajian:2014gza}. 

In this proceedings, we report initial results of using the Gamma-ray Burst Monitor (GBM) onboard the {\it Fermi} satellite to search for X-ray lines arising from sterile neutrino decay. Among the advantages of using the GBM for this purpose include: (i) its all-sky coverage, which allows the entire Milky Way dark matter halo to be studied, and (ii) the energy range of the GBM, which fills a gap in energy coverage between X-ray satellites and gamma-ray satellites. We therefore focus on the energy range $E_{\gamma} = 10$--$25$~keV corresponding to sterile neutrino mass $m_{s} = 20$--$50$~keV, and explore the Milky Way because of its proximity and well-studied dark matter distribution. 

\section{Expected signal}

\subsection{Intensity calculation}

The radiative decay of sterile neutrino has a decay rate of \citep{Pal:1981rm,Abazajian:2001vt},
\begin{equation}\label{eq:countrate}
\Gamma_s \simeq 1.36 \times 10^{-32} \, {\rm s^{-1}} \left( \frac{{\rm sin}^2 2\theta}{10^{-10}} \right) \left( \frac{m_s}{\rm 1 \, keV} \right)^5\,,
\end{equation}
where we have assumed a Majorana sterile neutrino.

\begin{table}[b]
\caption{\label{tab:table1}Dark matter profile parameters for widely adopted dark matter profiles in the literature. Our canonical profile is the NFW profile.}
\begin{tabular}{lccccc}
\hline
Profile		& $\alpha$	& $\beta$ 		& $\gamma$	& $R_s$ [pc]		\\
\hline
NFW			& $1$		& $3$ 		& $1$		& $20$			 	\\	
cNFW		& $1$		& $3$ 		& $1.15$		& $23.7$			 	\\
Isothermal	& $2$		& $2$ 		& $0$		& $3.5$		 		\\	
\end{tabular}
\end{table}

The photon intensity (number flux per solid angle) arising from sterile neutrino dark matter decay, from the direction of angle $\psi$ away from the Galactic Center, is
\begin{eqnarray}\label{eq:Intensity}
{\cal I}(\psi,E) &\equiv& \frac{dN}{dAdTd\Omega dE} \\
&=&  \frac{\rho_\odot R_\odot} {4\pi m_s \tau_s} \left(
 {\cal J}(\psi)\frac{dN}{dE} + R_{\rm EG} \int \frac{ dz}{h(z)}\frac{dN}{dE^{\prime}} \right) \nonumber \, ,
\end{eqnarray}
where the first term in the brackets is the contribution from the Galactic halo and the second term is the contribution from extragalactic halos. Here, $\tau_{s} = 1/\Gamma_s$ is the sterile neutrino lifetime, $\rho_\odot = 0.4 \, {\rm GeV \, cm^{-3}}$ is the local dark matter mass density, $R_\odot = 8.5$\,kpc is the Sun's distance to the galactic center, and $dN/dE = \delta(E - m_{s}/2)$ is the photon spectrum. ${\cal J}(\psi)$ is the so-called J-factor or boost factor, and is the integral of the dark matter mass density $\rho$ in the Milky Way halo along the line-of-sight,
\begin{equation}\label{eq:jfactor}
{\cal J}(\psi) = \frac{1}{\rho_\odot R_\odot} \int_{0}^{\ell_{max}} d\ell \; 
\rho(\psi,\ell) \, , 
\end{equation}
where $\ell_{max}$ is the outer limit of the dark matter halo. We assume the dark matter distribution is spherically symmetric about the Galactic Center, hence
\begin{eqnarray}
\rho(\psi,\ell)&=&\rho(r_{\rm GC}(\psi,\ell)) \\
&=& \rho\left(\sqrt{R_\odot^2-2\, \ell\, R_\odot\cos\psi+\ell^2} \right).
\end{eqnarray}
We adopt $\ell_{max}=250$\,kpc in this work. Although the value of $\ell_{max}$ differs depending on the adopted halo model, the contribution to ${\cal J}(\psi) $ from beyond $\sim 30$\,kpc are negligible. 

\begin{figure}[t]
\includegraphics[width=75mm]{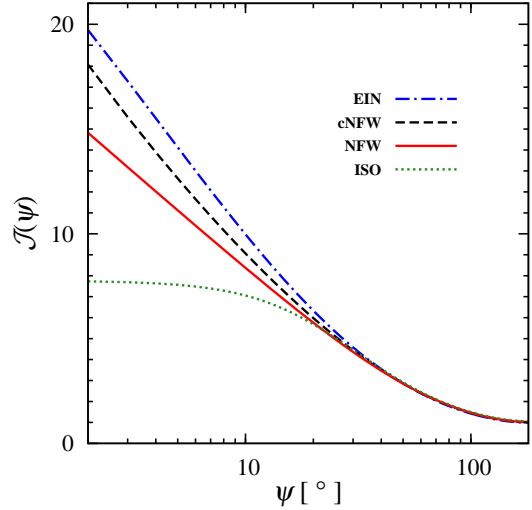}
\caption{\label{fig:Jfactor} The J-factor, $ {\cal J}(\psi)$, as a function of half opening angle $\psi$ relative to the Galactic Center, for four Milky Way dark matter halo profiles. Note the J-factor will be convolved by the GBM field of view, which is energy dependent but 40 degrees or more (see Figure \ref{fig:effa_angle}), which dramatically reduces the difference between the dark matter profiles for the analysis.}
\end{figure}

For the dark matter density profile $\rho$ we adopt a NFW profile as our canonical profile. Although the Milky Way dark matter density profile at small radii remains uncertain, it is known well enough for robust predictions of sterile neutrino decay signals on the scales of interest for the GBM. We adopt the following generic dark matter halo profile, motivated by numerical simulations,
\begin{equation}
\rho^{\alpha\beta\gamma} (r) = \rho_\odot \left( \frac{r}{R_\odot} \right)^{-\gamma} \left[ \frac{1+(R_\odot/R_s)^\alpha}{1+(r/R_s)^\alpha} \right]^{(\beta-\gamma)/\alpha},
\end{equation}
where parameters for commonly used profiles are summarized in Table \ref{tab:table1}. Another profile favored by recent simulations is the Einasto profile, 
\begin{equation}
\rho^{\rm Ein} (r) = \rho_\odot \, {\rm exp}\left( -\frac{2}{\alpha_E}\frac{r^{\alpha_E}-R_\odot^{\alpha_E}}{R_s^{\alpha_E}} \right),
\end{equation}
with $\alpha_E = 0.17$ and scale radius $R_s = 20$ kpc. The difference in the J-factor between these profiles are shown in Figure \ref{fig:Jfactor}. Here, the J-factors are shown as functions of the angle $\psi$ away from the Galactic Center. When the J-factor are convolved by the GBM field of view, which is energy dependent but typically $\sim 40$ degrees (see Figure \ref{fig:effa_angle}), the differences between dark matter profiles is dramatically reduced. 

\subsection{Signal modeling for the GBM}

The Fermi-GBM consists of $12$ NaI detectors and $2$ BGO detectors, the former covering energies 8 keV to 1 MeV, and the latter covering 200 keV to 40 MeV. The NaI detectors are physically placed on the corners and sides of the satellite. At any given time, 3 -- 4 NaI detectors view an Earth occultation, i.e., when the earth is within 60 degrees of the detector zenith. In the following, we perform a search using data from a single NaI detector, det-0. Det-0 is conveniently placed closest to the Fermi-LAT zenith (offset angle $20.6^{\circ}$). Thus, as the LAT engages in survey mode, which are designed to maximize sky coverage, so does det-0. Analyses on additional detectors are forthcoming. 

\begin{figure}[t]
\includegraphics[width=75mm]{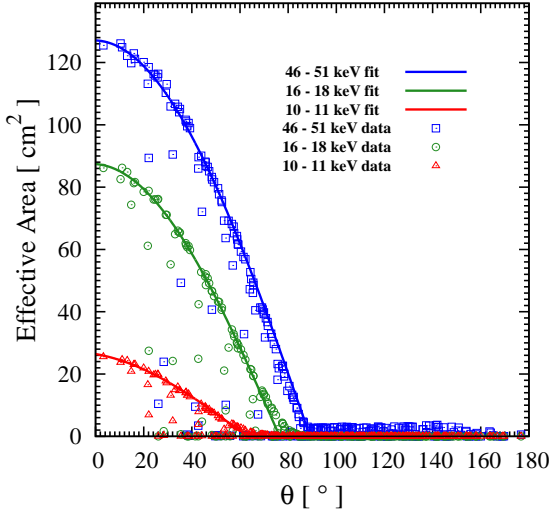} 
\caption{The effective area for det-0 NaI detector versus photon arrival angle with respect to the detector normal. Three energy ranges utilized in the sterile neutrino search are shown. The occasional dips in the effective area are due to blockages from the satellite components. \label{fig:effa_angle}}
\end{figure}

\begin{figure}[t] 
\includegraphics[width=75mm]{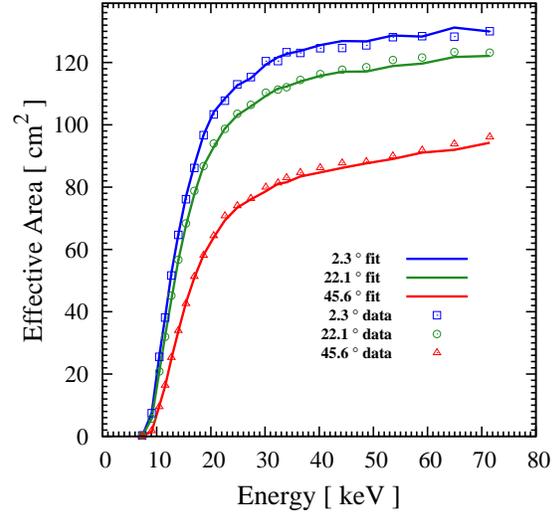} 
\caption{The same as Fig.~\ref{fig:effa_angle} but plotted against energy, for three incident angles. The fits come from the angular fits for various energies. \label{fig:effa_energy}}
\end{figure}

The response of det-0 are shown in Figures \ref{fig:effa_angle} and \ref{fig:effa_energy}, where the effective area is shown as functions of the angle away from the detector normal, $\theta$, and the photon energy. The angular dependence can be well-fitted with a Cosine function (shown by the solid line). The occasional dips are due to blockage from satellite components. 

One important consideration for the GBM is that it lacks photon-tracking capabilities, i.e., the photon count as a function of incident angle cannot be simply obtained. Earth occultation techniques can be effectively used to obtain directionality for point sources \cite{WilsonHodge:2012ix}, but remains difficult for diffuse sources. Fortunately, the lack of photon tracking ability is not very problematic for our sterile neutrino search, since the decay signal has a very large angular extent. However, it does mean it is difficult to accurately re-construct an intensity sky map (Eq.~\ref{eq:Intensity}). We opt to simulate the instrument observable by properly modeling the signal taking into account detector response. In this case, the instrumental observable is the photon count rate as a function of the NaI detector pointing direction.

The photon counts is energy dependent and direction dependent, i.e., $\nu_{i,j}$, where $i$ labels the energy bin and $j$ labels the detector pointing direction. The expected number of photons per observing time, $T_{j}$, from a particular detector pointing direction, is then 
\begin{eqnarray}
 \frac{d\nu_{i,j}}{dT_{j}} &=&\int_{E^{\rm min}_{i}}^{E^{\rm max}_{i}} dE\int_{2\pi} d\Omega(\theta) \int d\tilde{E}\, \\  
\nonumber && \times \Bigg\{ {\cal I}(\psi,\tilde{E}) \, G\left(E,\tilde{E}\right) \, A_{\rm eff}(\tilde{E}, \theta)   \Bigg\} \, , 
\end{eqnarray}
where $E^{\rm max}_{i}$ and $E^{\rm min}_{i}$ are the boundaries of the $i$-th energy bin.  We have integrated over the hemisphere the NaI detector points, i.e., over the detector zenith angle $\theta$, and attribute all the photons to the pixel defined by pointing $j$.  A position on the sky with an angle relative to the GC, $\psi$, is thus related to the detector zenith angle and the pixel that the detector points at through $\psi \rightarrow \psi(\theta, j)$\,.  The pointing direction of the detector is therefore defined by $\psi(0,j )$.  The factor $G(E,\tilde{E})$ takes into account the energy resolution of the NaI detector, which we model as a Gaussian with width given by the pre-launch calibrations \cite{Bissaldi:2008df, Meegan:2009qu}. Lastly,  $A_{\rm eff}(E, \theta)$ is the NaI detector effective area, which depends on energy and the detector zenith angle, as in Figures~\ref{fig:effa_angle}, \ref{fig:effa_energy}.

\section{Data reduction}

\begin{figure*} 
\begin{center}
\includegraphics[width=3.25in]{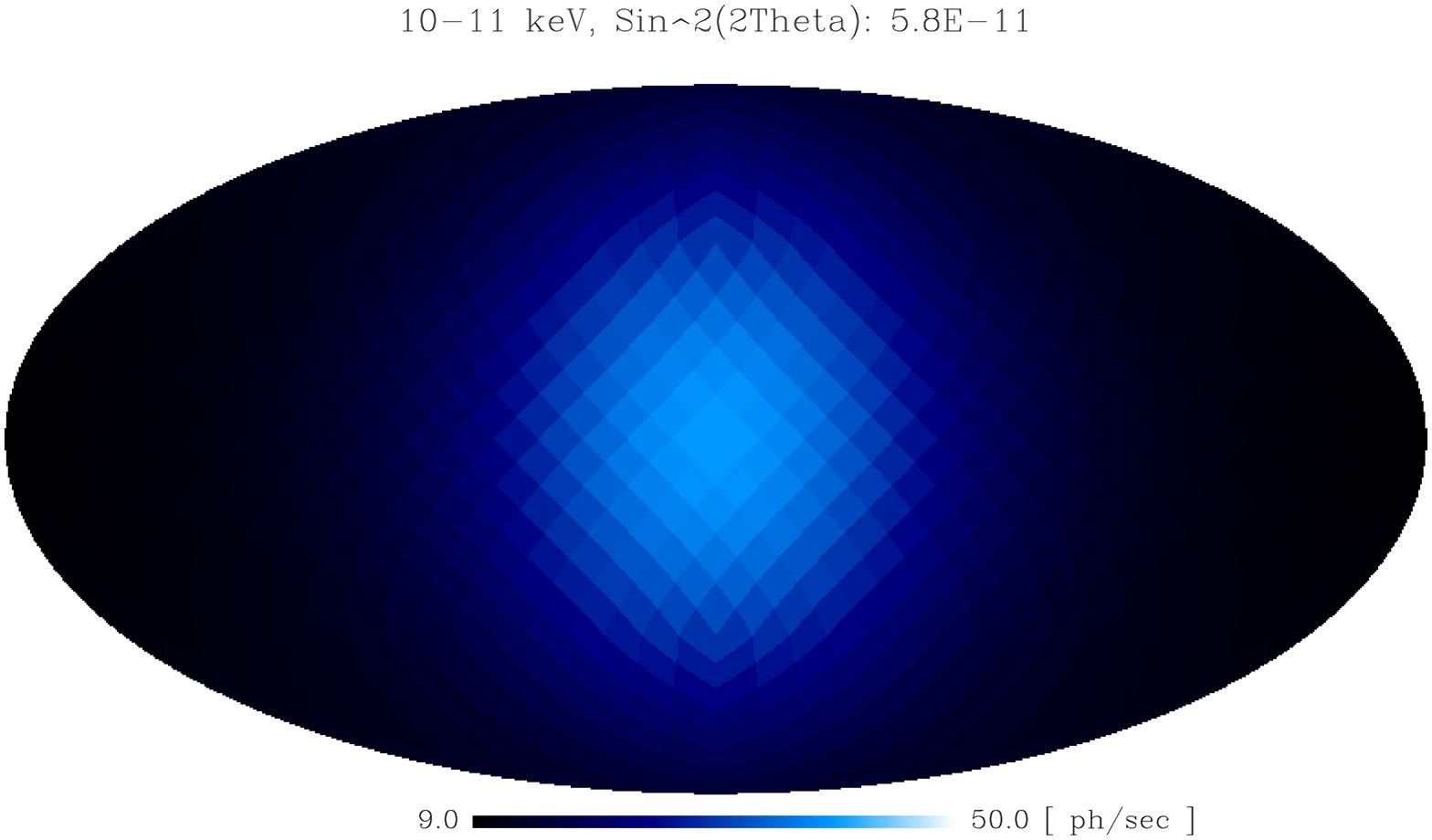} 
\includegraphics[width=3.25in]{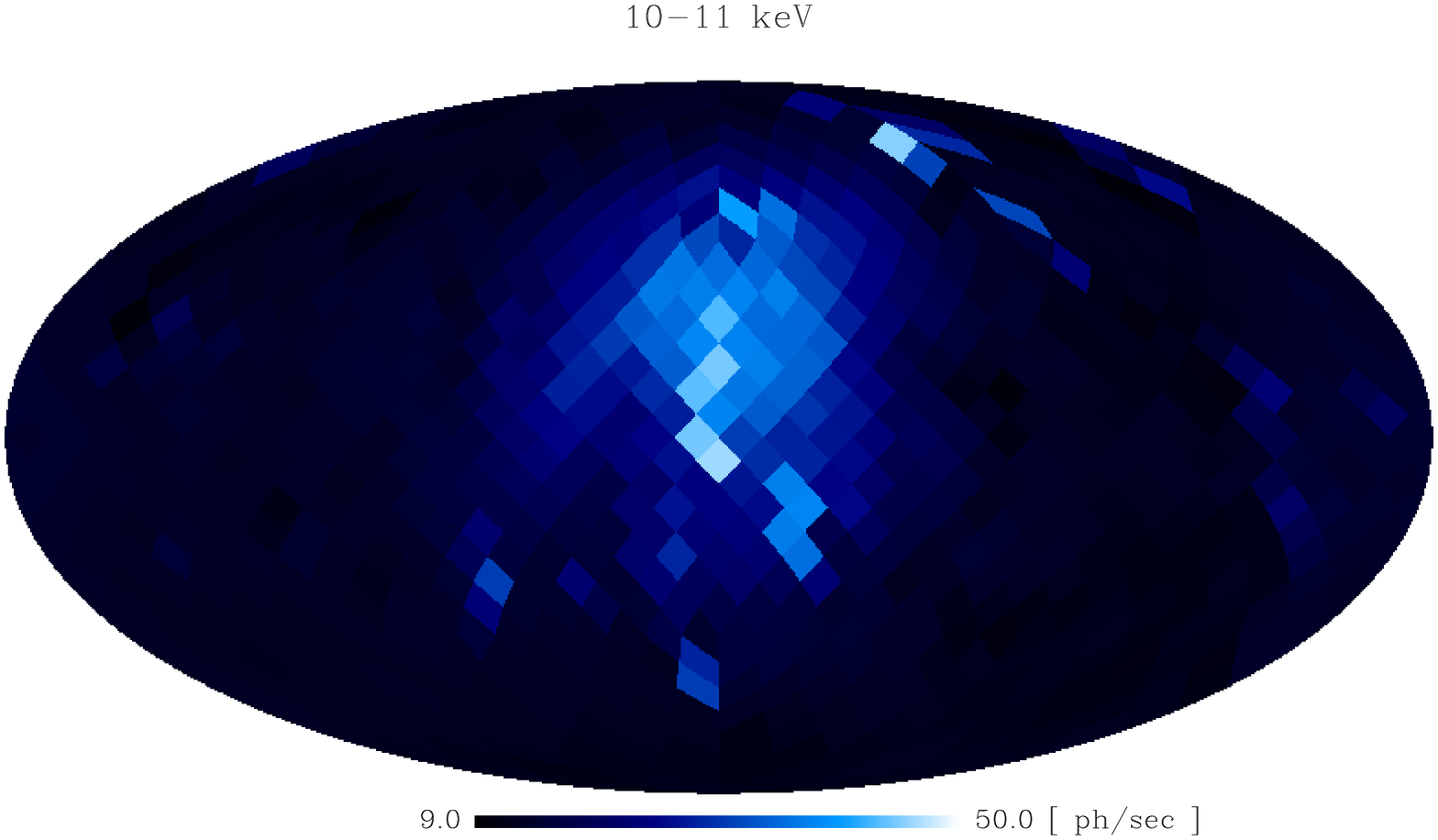} 
\caption{All-sky counts maps in 10 -- 11 keV energy range, showing the simulate dark matter map assuming the NFW profile (left) and the reduced data (right). The dark matter simulation assumes a sterile neutrino of mass 20 keV and mixing angle $\sin^{2}2\theta = 5.8 \times 10^{-11}$. For det-0 only. \label{fig:counts_rate_map} }
\end{center}
\end{figure*}

We use a total of 1601 days worth of data, from  12-AUG-2008 to 31-DEC-2012. We use the CSPEC data with nominal $4.096$\,s time resolution and $128$ channels in energy from 5 to 1402\,keV. We then implement a wide range of data cuts to minimize a wide range of background contributions, which are summarized below.

\begin{itemize}

\item LAT cuts: we use the Fermi-LAT cuts ``LAT\_CONFIG=1'', ``LAT\_MODE=5'', ``DATA\_QUAL=1'', ``ROCK\_ANGLE$<$50'', and ``SAA=F'', where the first three conditions ensure the detector configurations and data qualities are suitable for scientific analysis, and the fourth condition ensures the Earth is not in the LAT's field of view, which is approximately, although not exactly, the field of view of det-0 (this is addressed later). The cuts also exclude epochs where the satellite is passing through the South Atlantic Anomaly (SAA), which has high cosmic-ray activity that significantly increases the radioactivity of the satellite. 

\item Additional transient source cut: this removes epochs when the GBM detects transient sources, including gamma-ray bursts, direct cosmic-ray hits, solar flares, Galactic x-ray transients, and magnetospheric events. 

\item Extended SAA cut: the LAT cuts do not completely remove the effects of the SAA on the NaI detectors. The reason is that the satellite remains significantly radioactive even after leading the SAA. We therefore remove the data collected in orbits that pass through the SAA successively. 

\item Additional Earth cuts: we apply two additional cuts to remove backgrounds related to the Earth. The first requires the angle between the NaI detector normal and the vector from the Earth center to be less than $50^{\circ}$, which reduces emissions from the Earth limb. The second requires the geomagnetic latitude to be less than $|20|^{\circ}$, which reduces cosmic-ray induced vents.

\end{itemize}

The reduced data product contains observed counts in 128 energy bins and 768 equal area sky pixels in healpix projection. The total live time after reduction equals $\sim 4.6\times 10^{6}$ seconds ($\sim 53$ days). Despite the dramatic reduction in live time, the analysis is still systematic limited. In the energy range and region of interest, the total counts is more than $\sim 10^{7}$ photons for each energy bin. 

Figure \ref{fig:counts_rate_map} shows the simulated dark matter photon count rate (left) and the reduced data count rate (right) for det-0. Both are shown with the same dynamical range. Shown is the energy range 10--11 keV, which is our lower energy range. We observe a clear excess of photons towards the Galactic Center direction in the reduced data, which we interpret to be from astrophysical emissions from the Milky Way. Spectral analysis indicates the excess peaks at low energies, and is dramatically reduced by 30--40 keV, when the data is dominated by cosmic-ray induced backgrounds. This is confirmed by the high-energy sky map showing small variations that trace the Earth magnetic field structure. 

\section{Line search analysis}

Two line search strategies are implemented on the reduced data. The first is a conservative analysis based only on flux comparison, the second makes use of the spectral difference between signal (line) and background (dominated by a power-law within a small enough search energy window). 

\subsection{Flux analysis}

The most conservative constraint on a sterile neutrino decay amplitude is to require the decay signal counts do not exceed the total measured photon counts. Thus, we do not make any assumptions of the detector background and astrophysical background. 

The comparison is made bin by bin, and consists of comparing the predicted signal
\begin{eqnarray} \label{eq:pre_fs}
\nu_i &=& \sum_{j}^{ROI} T_{j} \frac{d\nu_{i,j}}{dT_{j}} \, ,
\end{eqnarray}
to the observed photon counts, 
\begin{equation} \label{eq:data}
d_i =  \sum_{j}^{ROI} N_{i,j}  \,,
\end{equation}
where $N_{i,j}$ is the number of counts in energy bin $i$ and pixel $j$ measured by the det-0. 

\subsection{Spectral analysis}

The flux analysis can be improved by modeling the background. We perform a spectral analysis which captures the different spectral shapes of the sterile neutrino decay signal and the backgrounds. 

The background is modeled as a power-law,
\begin{eqnarray}
\frac{db}{dE} &=&  \beta \left( \frac{E}{E_0} \right)^{-\gamma} {\cal N}(E) \, ,   \label{eq:bkg}
\end{eqnarray}
where the normalization $\beta$ and index $\gamma$ are left as free parameters. This assumption is made in a small energy window defined by
\begin{equation} \label{eq:Ewin}
{\rm Max} \left( i_{\rm min}, i-\Delta w  \right)< i < i +\Delta w , 
\end{equation}
using a fixed window of $\Delta w = 5$. This choice results in each energy width side being about $3-4\,\sigma$ of the energy resolution. The energy window is truncated at $i_{\rm min}= 6$, which corresponds to a central bin energy of 9.4~keV. For line energies near the low energy cutoff, the energy window is thus asymmetric. 

\begin{figure}[t] 
\includegraphics[width=75mm]{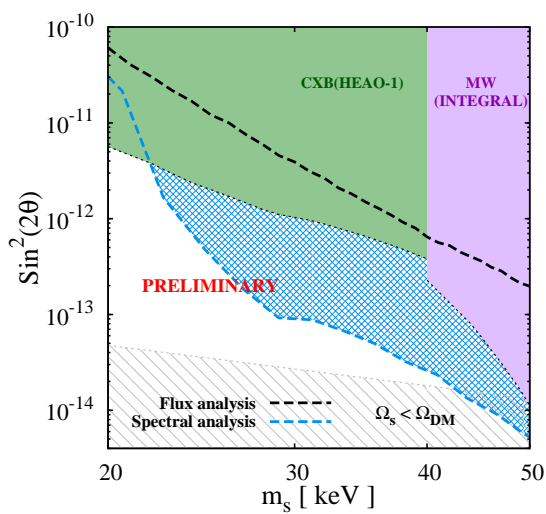} 
\caption{Constraints on the sterile neutrino mixing angle $\sin^{2}(2\theta)$ as a function of the sterile neutrino mass $m_s$. Our new limits are shown by the dashed lines: the upper (black) line is from our conservative flux limit analysis, while the stronger lower (blue) line is from modeling the background as a power-law. The latter limit improves by preceding ones, obtained by observations of cosmic X-ray background (CXB) by HEAO-1~\cite{Boyarsky:2006fg}, by about an order of magnitude.}
\label{fig:constraint}
\end{figure}

We first perform a $\chi^{2}$ test to assess whether the assumption of a power-law is a good local background model. By way of minimizing the negative logarithm of the likelihood function, we find the best-fit $\beta_0$ and $\gamma_0$ for all energy windows defined by Eq.~(\ref{eq:Ewin}). We assume a Gaussian probability distribution function for each energy bin, giving the likelihood function as
\begin{equation}
{\cal L}(\beta,\gamma | d) = \prod_{i} \frac{1}{\sqrt{2 \pi} \sigma_{\rm Aeff} } e^{-\frac{{ (b_{i}-d_{i})^{2}}}{2 {\sigma_{\rm Aeff}}^{2} } } \, ,
\end{equation}
where the product is taken over the energy bins, and $\sigma_{\rm Aeff}$ is the systematic uncertainty of the effective area, 
\begin{equation}
\sigma_{\rm Aeff}  = 0.05 (\nu_{i} + b_{i})  \, .
\end{equation}
The present analysis is not limited by statistical uncertainty; rather, it is dominated by systematic uncertainties. We adopt a constant 5\% uncertainty on the effective area as a conservative choice. The $\chi^{2}$ per degree of freedom for most energy windows is found to cluster between 1.1 and 1.4, although in some higher energy cases they can be as low as 0.5. We conclude that these findings justify the use of the local power-law assumption, as well as the 5\% uncertainty in the effective area. 

Finally, the sterile neutrino decay line is added to the central energy bin for each energy window. The line signal has one free parameter, $f_s$, the normalization, which scales linearly with the mixing angle ${\rm sin}^2 2 \theta$. Thus, the search has three free parameters, $\beta$, $\gamma$, and $f_s$. The former two are treated as nuisance parameters, and $f_s$ is the parameter of interest. The likelihood is
\begin{equation}
{\cal L}(f_{s}, { \kappa} | d) = \prod_{i} \frac{1}{\sqrt{2 \pi} \sigma_{\rm Aeff} } e^{-\frac{{ (\nu_{i} + b_{i}-d_{i})^{2}}}{2 {\sigma_{\rm Aeff}}^{2} } } \, ,
\end{equation}
where $\kappa$ are the nuisance parameters. Finally, we use the profile likelihood analysis \cite{Rolke:2004mj} for treating nuisance parameters. In practice, this involves calculating the profile likelihood $- {\rm ln} {\cal L}_p (f_s) $ for several fixed values of $m_s$, where for each $f_s$ the $-{\rm ln} {\cal L}$ is minimized with respect to all other parameters $\kappa$. 

\section{Sterile neutrino limits}

We find no significant detections of lines in both our analyses. We determine the 95\% C.L. one-sided upper limits on the amplitude, $f_s^{95}$, by requiring a $2 \Delta {\rm ln} {\cal L} = 2.71$.  The limits are shown in Figure \ref{fig:constraint}. While the limits from the conservative flux analysis is weaker than previous limits, our spectral analysis limit results in an improvement of about a factor of $\sim 10$ compared to those based on observations of cosmic X-ray background (CXB) by HEAO-1~\cite{Boyarsky:2006fg}. The spectral analysis limit deteriorates at low energy, because the number of energy bins decreases, and also because the energy window is increasingly asymmetric. The latter in particular results in reduced ability to distinguish between the power-law background model and the line signal. 

\section{Conclusions and discussions}

We have used a GBM NaI detector (det-0) to set upper limits on sterile neutrino dark matter decays into mono-energetic photons. Two line analyses were implemented: the first is a conservative flux search, and the second is a spectral search assuming the background can be modeled as a power-law over small energy windows. In the energy range of 10--25 keV, corresponding to sterile neutrino mass of 20--50 keV, our new upper limit is an improvement of about an order of magnitude compared to previous limits using the CXB with {\it HEAO-1} data. 

The current analyses are dominated by systematic uncertainties primarily in the effective area. A better understanding of the GBM detector will therefore improve the limits presented in this proceeding. Work is currently underway to investigate other NaI detectors onboard the GBM.

\bigskip 
\begin{acknowledgments}
We thank John Beacom for useful discussions. We also thank the Fermi-GBM team for providing the transient event time intervals. K.C.Y.N. was supported by a Fermi Guest Investigator award (Cycle 4), and NSF Grant PHY-1101216 (to John Beacom).
\end{acknowledgments}

\bigskip 

\begin{thebibliography}{99} 

\bibitem{Kusenko:2009up} 
  A.~Kusenko,
  Phys.\ Rept.\  {\bf 481}, 1 (2009)
  [arXiv:0906.2968 [hep-ph]].
\bibitem{Boyarsky:2009ix} 
  A.~Boyarsky, O.~Ruchayskiy and M.~Shaposhnikov,
  Ann.\ Rev.\ Nucl.\ Part.\ Sci.\  {\bf 59}, 191 (2009)
  [arXiv:0901.0011 [hep-ph]].
  \bibitem{Dodelson:1993je} 
  S.~Dodelson and L.~M.~Widrow,
  Phys.\ Rev.\ Lett.\  {\bf 72}, 17 (1994)
  [hep-ph/9303287].
\bibitem{Shi:1998km} 
  X.~D.~Shi and G.~M.~Fuller,
  Phys.\ Rev.\ Lett.\  {\bf 82}, 2832 (1999)
  [astro-ph/9810076].
\bibitem{Abazajian:2001nj} 
  K.~Abazajian, G.~M.~Fuller and M.~Patel,
  Phys.\ Rev.\ D {\bf 64}, 023501 (2001)
  [astro-ph/0101524].  
\bibitem{Asaka:2006nq} 
  T.~Asaka, M.~Laine and M.~Shaposhnikov,
  JHEP {\bf 0701}, 091 (2007)
  [hep-ph/0612182].  
\bibitem{Boyarsky:2008xj} 
  A.~Boyarsky, J.~Lesgourgues, O.~Ruchayskiy and M.~Viel,
  JCAP {\bf 0905}, 012 (2009)
  [arXiv:0812.0010 [astro-ph]].  
\bibitem{Boyanovsky:2010pw} 
  D.~Boyanovsky and J.~Wu,
  Phys.\ Rev.\ D {\bf 83}, 043524 (2011)
  [arXiv:1008.0992 [astro-ph.CO]].  
\bibitem{Kusenko:1997sp} 
  A.~Kusenko and G.~Segre,
  Phys.\ Lett.\ B {\bf 396}, 197 (1997)
  [hep-ph/9701311].  
\bibitem{Hidaka:2006sg} 
  J.~Hidaka and G.~M.~Fuller,
  Phys.\ Rev.\ D {\bf 74}, 125015 (2006)
  [astro-ph/0609425].
\bibitem{Fuller:2003gy} 
  G.~M.~Fuller, A.~Kusenko, I.~Mocioiu and S.~Pascoli,
  Phys.\ Rev.\ D {\bf 68}, 103002 (2003)
  [astro-ph/0307267].
\bibitem{Barkovich:2004jp} 
  M.~Barkovich, J.~C.~D'Olivo and R.~Montemayor,
  Phys.\ Rev.\ D {\bf 70}, 043005 (2004)
  [hep-ph/0402259].
\bibitem{Fryer:2005sz} 
  C.~L.~Fryer and A.~Kusenko,
  Astrophys.\ J.\ Suppl.\  {\bf 163}, 335 (2006)
  [astro-ph/0512033].
\bibitem{Abazajian:2001vt} 
  K.~Abazajian, G.~M.~Fuller and W.~H.~Tucker,
  Astrophys.\ J.\  {\bf 562}, 593 (2001)
  [astro-ph/0106002].
\bibitem{Bulbul:2014sua} 
  E.~Bulbul, M.~Markevitch, A.~Foster, R.~K.~Smith, M.~Loewenstein and S.~W.~Randall,
  Astrophys.\ J.\  {\bf 789}, 13 (2014)
  [arXiv:1402.2301 [astro-ph.CO]].
\bibitem{Boyarsky:2014jta} 
  A.~Boyarsky, O.~Ruchayskiy, D.~Iakubovskyi and J.~Franse,
  Phys.\ Rev.\ Lett.\  {\bf 113}, no. 25, 251301 (2014)
  [arXiv:1402.4119 [astro-ph.CO]].
\bibitem{Riemer-Sorensen:2014yda} 
  S.~Riemer-Sorensen,
  arXiv:1405.7943 [astro-ph.CO].
\bibitem{Jeltema:2014qfa} 
  T.~E.~Jeltema and S.~Profumo,
  arXiv:1408.1699 [astro-ph.HE].
\bibitem{Boyarsky:2014ska} 
  A.~Boyarsky, J.~Franse, D.~Iakubovskyi and O.~Ruchayskiy,
  arXiv:1408.2503 [astro-ph.CO].
\bibitem{Malyshev:2014xqa} 
  D.~Malyshev, A.~Neronov and D.~Eckert,
  Phys.\ Rev.\ D {\bf 90}, no. 10, 103506 (2014)
  [arXiv:1408.3531 [astro-ph.HE]].
\bibitem{Anderson:2014tza} 
  M.~E.~Anderson, E.~Churazov and J.~N.~Bregman,
  arXiv:1408.4115 [astro-ph.HE].
\bibitem{Tamura:2014mta} 
  T.~Tamura, R.~Iizuka, Y.~Maeda, K.~Mitsuda and N.~Y.~Yamasaki,
  arXiv:1412.1869 [astro-ph.HE].
\bibitem{Urban:2014yda} 
  O.~Urban, N.~Werner, S.~W.~Allen, A.~Simionescu, J.~S.~Kaastra and L.~E.~Strigari,
  arXiv:1411.0050 [astro-ph.CO].
\bibitem{Abazajian:2014gza} 
  K.~N.~Abazajian,
  Phys.\ Rev.\ Lett.\  {\bf 112}, no. 16, 161303 (2014)
  [arXiv:1403.0954 [astro-ph.CO]].
\bibitem{Pal:1981rm} 
  P.~B.~Pal and L.~Wolfenstein,
  Phys.\ Rev.\ D {\bf 25}, 766 (1982).
\bibitem{WilsonHodge:2012ix} 
  C.~A.~Wilson-Hodge, G.~L.~Case, M.~L.~Cherry, J.~Rodi, A.~Camero-Arranz, P.~Jenke, V.~Chaplin and E.~Beklen {\it et al.},
  Astrophys.\ J.\ Suppl.\  {\bf 201}, 33 (2012)
  [arXiv:1201.3585 [astro-ph.HE]].
\bibitem{Bissaldi:2008df} 
  E.~Bissaldi, A.~von Kienlin, G.~Lichti, H.~Steinle, P.~N.~Bhat, M.~S.~Briggs, G.~J.~Fishman and A.~S.~Hoover {\it et al.},
  Exper.\ Astron.\  {\bf 24}, 47 (2009)
  [arXiv:0812.2908 [astro-ph]].
\bibitem{Meegan:2009qu} 
  C.~Meegan, G.~Lichti, P.~N.~Bhat, E.~Bissaldi, M.~S.~Briggs, V.~Connaughton, R.~Diehl and G.~Fishman {\it et al.},
  Astrophys.\ J.\  {\bf 702}, 791 (2009)
  [arXiv:0908.0450 [astro-ph.IM]].
\bibitem{Rolke:2004mj} 
  W.~A.~Rolke, A.~M.~Lopez and J.~Conrad,
  Nucl.\ Instrum.\ Meth.\ A {\bf 551}, 493 (2005)
  [physics/0403059].
\bibitem{Boyarsky:2006fg} 
  A.~Boyarsky, A.~Neronov, O.~Ruchayskiy, M.~Shaposhnikov and I.~Tkachev,
  Phys.\ Rev.\ Lett.\  {\bf 97}, 261302 (2006)
  [astro-ph/0603660].


\end{thebibliography}

\end{document}